\documentclass[pra,showpacs,twocolumn,amsmath,amssymb]{revtex4}
\usepackage{graphicx}
\begin{document}
\title{Purification and detection of entangled coherent states}
\date{\today}
\author{J. Clausen}
\email{J.Clausen@tpi.uni-jena.de}
\author{L. Kn\"oll}
\author{D.-G. Welsch}
\affiliation{Friedrich-Schiller-Universit\"at Jena,\\
             Theoretisch-Physikalisches Institut,\\
             Max-Wien-Platz 1, D-07743 Jena, Germany}
\begin{abstract}
In [J. C. Howell and J. A. Yeazell, Phys. Rev. A \textbf{62}, 012102 (2000)], a
proposal is made to generate entangled macroscopically distinguishable states
of two spatially separated traveling optical modes. We model the
decoherence due to light scattering during the propagation along an optical
transmission line and propose a setup allowing an entanglement purification
from a number of preparations which are partially decohered due to
transmission. A purification is achieved even without any manual intervention.
We consider a nondemolition configuration to measure the purity of the state as
contrast of interference fringes in a double-slit setup. Regarding the
entangled coherent states as a state of a bipartite quantum system, a close
relationship between purity and entanglement of formation can be obtained. In
this way, the contrast of interference fringes provides a direct means to
measure entanglement.
\end{abstract}
\maketitle
\section{\label{sec1}
       Introduction}
The preparation of two spatially separated traveling optical modes in
entangled coherent states is of special interest since by representing an
outcome of Schr\"odinger's thought experiment \cite{catSchrodinger} and at the
same time a state of a bipartite quantum system \cite{entangledBennett}, it
provides a link between philosopical underpinnings of quantum mechanics and
applications in quantum information processing. On the other hand, this state
has a chance of its experimental realization in the laboratory.

In this paper, we continue the study of two-mode entangled coherent states,
whose preparation is investigated in \cite{catHowell}. The setup considered
in \cite{catHowell} applies a Mach-Zehnder interferometer equipped with a
cross-Kerr element in each of two spatially separated modes. In turn, it may be
regarded as a two-mode extension of an analogous single-mode configuration that
is proposed in \cite{catGerry} to prepare a superposition of two single-mode
coherent states. Here, we put emphasis on the state transmission,
purification and detection as subsequent steps of quantum state engineering.
In doing so, these schemes can be linked to other work on detection and
application of single- and two-mode quantum states. For example, a double-slit
configuration in combination with a cross-Kerr element is used in
\cite{slitFilip} to measure in terms of interference fringe contrast a variety
of quantities representing measures of quantum state distance. Moreover, there
is a number of works treating the properties of general interferometry with
entangled coherent states \cite{coh1Rice,coh2Rice,coh3Rice,coh4Rice,coh5Rice}. 
Applications to quantum teleportation of one qubit are considered in
\cite{teleLee} for the example of entangled single-photon states and in
\cite{catEnk} for the example of entangled coherent states. In \cite{teleLee} also
Bell's inequality is studied and in \cite{catEnk} the effect of decoherence is
analysed. A quantum nondemolition setup to detect presence of a single photon
is proposed in \cite{triggerHowell}.

In the present work, we pay special attention to entanglement purification,
which plays an important role in quantum information processing
\cite{entangledBennett} and becomes a delicate issue especially if entangled
macroscopically distinguishable quantum states are involved. In particular, we
address the question of state evolution if a setup implementing an entanglement
purification protocol is left on its own. It turns out that the state purifies
itself, albeit with random parity.

The state preparation is achieved by a conditional state reduction connected
with a measurement. With the consideration of a spatially extented
configuration, the concept of entanglement comes into play. Entanglement cannot
be observed in a local context. Therefore, for a such a conditional scheme to
be meaningful, there should be an observer who, like a viewer of the drawings
in the figures below, has an overlook of the actions performed elsewhere and
their results. In our setups, this will be a third party located halfway
between the two entangled parties and which communicates with them either by a
classical or a quantum channel. Below, we will see that either channel type may
be used, depending on the experimental situation.
\section{\label{sec2}
       State preparation and transmission}
\begin{figure}[ht]
\includegraphics[width=8.6cm]{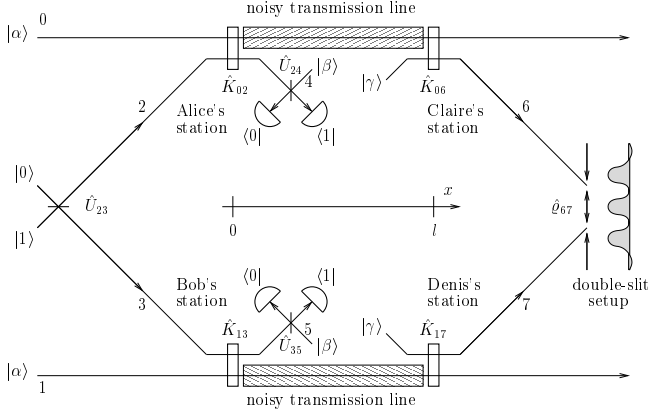}
\caption{\label{fig1}
Preparation, transmission and detection of entangled mesoscopic coherent
states. The entanglement remaining after having been subjected to scattering
loss during transmission is measured as the contrast of interference fringes
seen in a double-slit setup.
}
\end{figure}

The scheme relies on cross-Kerr couplers described by operators
$\hat{K}_{jk}$ $\!=$ $\!\mathrm{e}^{\mathrm{i}\pi
\hat{n}_j\hat{n}_k}$, where $\hat{n}_j$ $\!=$ $\hat{a}_j^\dagger\hat{a}_j$, and
beam splitters described by operators $\hat{U}_{jk}(T,R)$ defined via the
relation
\begin{equation}
  \hat{U}_{jk}^{\dagger}\binom{\hat{a}_j}{\hat{a}_k}\hat{U}_{jk}
  =\left(\begin{array}{cc}
  T & R \\
  -R^* & T^*
  \end{array}\right)
  \binom{\hat{a}_j}{\hat{a}_k}
\label{MC}
\end{equation}
by their complex transmittance $T$ and reflectance $R$ obeying
$|T|^2$ $\!+$ $\!|R|^2$ $\!=$ $\!1$. We start with the state preparation.
Fig.~\ref{fig1} shows the overall setup.
The source of the entangled state consists of a device able to prepare a pulse
in a single-photon state $|1\rangle$, which is fed into one of the input ports
of a balanced beam splitter $\hat{U}_{23}(T=-R=1/\sqrt{2})$ whose second input
port remains in the vacuum state $|0\rangle$. The pulse leaving this beam
splitter is therefore prepared in an entangled two-mode state
\begin{equation}
  \hat{U}_{23}|1\rangle_2|0\rangle_3
  =\frac{|0\rangle_2|1\rangle_3+|1\rangle_2|0\rangle_3}{\sqrt{2}},
\label{s}
\end{equation}
with a mode 2 traveling to Alice's and a mode 3 traveling to Bob's station.
Alice and Bob use cross-Kerr couplers $\hat{K}_{02}$ and $\hat{K}_{13}$ to mix
the received pulse with an incoming signal pulse corresponding to their signal
mode 0 and 1, respectively. After that, each of them mixes the pulse from
$\hat{U}_{23}$ with one prepared in a coherent state
$|\beta\rangle$ $\!=$ $\!\exp(-|\beta|^2/2)\sum_k\beta^k/\sqrt{k!}|k\rangle$
(mode 4 and 5) using identical beam splitters $\hat{U}_{24}(T,R)$ and
$\hat{U}_{35}(T,R)$, respectively, and performs photon number measurements.

If Alice and Bob detect with their photodetectors 0 and 1 photons as depicted
in Fig.~\ref{fig1}, their combined action on the signal state can be described
by a (\textquotedblleft{conditional}\textquotedblright\,) two-mode operator 
\begin{eqnarray}
  \hat{Y}
  &=&\,_{2}\langle1|\,_{3}\langle1|\,_{4}\langle0|\,_{5}\langle0|
  \hat{U}_{35}\hat{U}_{24}\hat{K}_{13}\hat{K}_{02}
  |\beta\rangle_{4}|\beta\rangle_{5}\hat{U}_{23}|1\rangle_2|0\rangle_3
  \nonumber\\
  &=&\frac{TR\langle0|\beta\rangle\langle1|\beta\rangle}
  {\sqrt{2}}[(-1)^{\hat{n}_0}+(-1)^{\hat{n}_1}].
\label{y}
\end{eqnarray}
Note that with regard to the detection result as shown in Fig.~\ref{fig1},
Alice's photodetectors together with the beam splitter $\hat{U}_{24}$ and the
coherent state input may be regarded as a unit performing a detection
\begin{equation}
  \,_2\langle1|\,_4\langle0|\hat{U}_{24}|\beta\rangle_4
  =\mathrm{e}^{-\frac{|\beta|^2}{2}}
  \Bigl(\,_2\langle1|\,T+\,_2\langle0|\,R\beta\Bigr)
\end{equation}
in mode 2 (the same holds for the corresponding unit on Bob's side).
When Alice's and Bob's signal pulses are initially prepared in coherent states
$|\alpha\rangle$, the two-mode state of the signal pulses leaving $\hat{K}_{02}$
and $\hat{K}_{13}$ becomes
\begin{eqnarray}
  |\Psi\rangle_{01}&=&\frac{1}{\sqrt{p}}\hat{Y}|\alpha\rangle_0|\alpha\rangle_1
  \nonumber\\&=&
  \frac{|\alpha\rangle_0|\!-\!\alpha\rangle_1
  +|\!-\!\alpha\rangle_0|\alpha\rangle_1}
  {\sqrt{2(1+\mathrm{e}^{-4|\alpha|^2})}}.
\label{st}
\end{eqnarray}
The probability
\begin{eqnarray}
  p&=&\,_{0}\langle\alpha|\,_{1}\langle\alpha|\hat{Y}^\dagger
  \hat{Y}|\alpha\rangle_0|\alpha\rangle_1
  \nonumber\\&=&
  |TR\langle0|\beta\rangle\langle1|\beta\rangle|^2(1+\mathrm{e}^{-4|\alpha|^2})
\end{eqnarray}
takes for $|T|^2$ $\!=$ $\!|\beta|^2$ $\!=$ $\!1/2$ the maximal value
\begin{equation}
  p_\mathrm{max}
  =\frac{1+\mathrm{e}^{-4|\alpha|^2}}{8\mathrm{e}}
  \approx(8\mathrm{e})^{-1}\approx5\%,
\label{pmax}
\end{equation}
which becomes for sufficiently large $|\alpha|$ independent of $\alpha$.
Finally, the classical trigger signal confirming the desired photodetection
result is sent to an (observing) third party.

We see that Eq.~(\ref{st}) is a superposition of macroscopically
distinguishable states $|\alpha\rangle_0|\!-\!\alpha\rangle_1$ and
$|\!-\!\alpha\rangle_0|\alpha\rangle_1$. This means that the two-mode
entanglement of a \textit{single photon} triggers the preparation of entangled
coherent states with \textit{arbitrarily large} amplitude $|\alpha|$. With regard
to entanglement, the case $|\alpha|$ $\!\gg$ $\!1$ describes the situation of
two Schr\"odinger's cats \cite{catSchrodinger}: The entangled single-photon state
Eq.~(\ref{s}) represents the (microscopic) radioactive atom, the devices at
Alice's and Bob's station represent the (amplifying) mechanisms releasing the
poison, and the signal pulse whose state changes from the initial product state
$|\alpha\rangle_0|\alpha\rangle_1$ to Eq.~(\ref{st}) represents two (macroscopic)
cats, one of which is killed but it is uncertain which one.
\begin{itemize}
\item \textit{Alternative:}
\end{itemize}
The state Eq.~(\ref{st}) can also be generated if Alice and Bob prepare the
second input modes of $\hat{K}_{02}$ and $\hat{K}_{13}$ in coherent states
$|\beta\rangle$ and send the respective output pulses to a third location where
they are mixed using a beam splitter $\hat{U}_{23}^\dagger$. If behind this
beam splitter 1 and 0 photons are detected in mode 2 and 3, respectively, their
action can again be described by a conditional operator
\begin{eqnarray}
  \hat{\tilde{Y}}
  &=&\,_2\langle1|\,_3\langle0|
  \hat{U}_{23}^\dagger\hat{K}_{13}\hat{K}_{02}
  |\beta\rangle_2|\beta\rangle_3
  \nonumber\\
  &=&\frac{\langle0|\beta\rangle\langle1|\beta\rangle}
  {\sqrt{2}}[(-1)^{\hat{n}_0}+(-1)^{\hat{n}_1}],
\label{ytilde}
\end{eqnarray}
and the signal state reads
\begin{equation}
  \frac{1}{\sqrt{\tilde{p}}}\hat{\tilde{Y}}|\alpha\rangle_0|\alpha\rangle_1
  =|\Psi\rangle_{01}.
\end{equation}
For $\beta$ $\!=$ $\!1/\sqrt{2}$ the success probability becomes
$\tilde{p}$ $\!=$ $\!4p_\mathrm{max}$, compare Eq.~(\ref{st}) and Eq.~(\ref{pmax}).
The result of this measurement does of course not affect the reduced signal
state
\begin{eqnarray}
  &&\mathrm{Tr}_2\Bigl(\hat{K}_{02}|\alpha\rangle_0|\beta\rangle_2
  \,_0\langle\alpha|\,_2\langle\beta|\hat{K}_{02}^\dagger\Bigr)
  \nonumber\\
  &&=C(|\beta|^2)|\alpha\rangle_0\langle\alpha|
  +S(|\beta|^2)|\!-\!\alpha\rangle_0\langle-\alpha|
\label{rs}
\end{eqnarray}
as observed by Alice (analogously Bob) locally. It only changes the information
available at the third location. In Eq.~(\ref{rs}) we have used the functions
\begin{subequations}
\begin{eqnarray}
  C(x)&=&\mathrm{e}^{-x}\cosh x,\\
  S(x)&=&\mathrm{e}^{-x}\sinh x.
\end{eqnarray}
\label{cs-a}
\end{subequations}

Let us now consider the transmission of a signal in a two-mode state
$\hat{\varrho}(x)$ along a lossy transmission line ranging from
$x$ $\!=$ $\!0$ to $x$ $\!=$ $\!l$ (cf. Fig.~\ref{fig1}).
In order to model the scattering loss occuring during transmission, we insert
at locations $x_k$ $\!=$ $\!kL/n$, $k$ $\!=$ $\!0,1,\ldots$ beam splitters
$\hat{U}_{0k}$ with transmittances $\!T_{0}^{1/n}$ in Alice's signal mode and
beam splitters $\hat{U}_{1k}$ with transmittances $\!T_{1}^{1/n}$ in Bob's
signal mode \cite{slitWeb}. Here, $T_{0/1}$ $\!\in$ $\![0,1]$ define the
scattering losses per given length $L$. If all the second input modes of these
beam splitters are in the vacuum state, we obtain the relation
\begin{eqnarray}
  \hat{\varrho}(x+L/n)&=&\mathrm{Tr}_{kk^\prime}\left[
  \hat{U}_{1k^\prime}\hat{U}_{0k}
  |0\rangle_{kk^\prime}\hat{\varrho}(x)\,_{kk^\prime}\langle0|
  \hat{U}_{0k}^\dagger\hat{U}_{1k^\prime}^\dagger
  \right]
  \nonumber\\
  &=&\sum_{l,l^\prime=0}^\infty
  \frac{(T_0^{-2/n}-1)^l(T_1^{-2/n}-1)^{l^\prime}}{l!l^\prime!}
  \nonumber\\
  &&\times\hat{a}_1^{l^\prime}\hat{a}_0^l
  T_1^{\frac{\hat{n}_1}{n}}T_0^{\frac{\hat{n}_0}{n}}
  \hat{\varrho}(x)
  T_0^{\frac{\hat{n}_0}{n}}T_1^{\frac{\hat{n}_1}{n}}
  \hat{a}_0^{\dagger\,l}\hat{a}_1^{\dagger\,l^\prime}.
\label{fe}
\end{eqnarray}
By expanding the exponentials and replacing differences with differentials one
may verify that in the limit $n$ $\!\rightarrow$ $\!\infty$, Eq.~(\ref{fe}) yields
the well-known master equation of two damped harmonic oscillators
\cite{masterWalls}
\begin{equation}
  L\frac{\mathrm{d}\hat{\varrho}(x)}{\mathrm{d}x}
  =\ln T_{0}\mathcal{L}_0\hat{\varrho}(x)
  +\ln T_{1}\mathcal{L}_1\hat{\varrho}(x),
\label{me}
\end{equation}
where
\begin{equation}
  \mathcal{L}_j\hat{\varrho}
  =\hat{n}_j\hat{\varrho}
  -2\hat{a}_j\hat{\varrho}\hat{a}_j^\dagger
  +\hat{\varrho}\hat{n}_j.
\end{equation}
(Note that the unitary state evolution describing the pulse propagation in
free space is taken into account in the interaction picture, which may be
envisaged as changing to a co-rotating frame in phase space.)
It is sufficient to limit attention to states of the form
\begin{equation}
  \hat{\varrho}(x)=\frac{\hat{\rho}_\mathrm{inc}+r\hat{\rho}_\mathrm{coh}}
  {2[1+r\mathrm{e}^{-2(|\alpha_0|^2+|\alpha_1|^2)}]},
\label{initstate}
\end{equation}
where
\begin{eqnarray*}
  \hat{\rho}_\mathrm{inc}
  &=&|\alpha_0,-\alpha_1\rangle\langle\alpha_0,-\alpha_1|
  +|\!-\!\alpha_0,\alpha_1\rangle\langle-\alpha_0,\alpha_1|,
  \\
  \hat{\rho}_\mathrm{coh}
  &=&|\alpha_0,-\alpha_1\rangle\langle-\alpha_0,\alpha_1|
  +|\!-\!\alpha_0,\alpha_1\rangle\langle\alpha_0,-\alpha_1|
\end{eqnarray*}
($|\alpha,\beta\rangle$ $\!\equiv$ $\!|\alpha\rangle_0|\beta\rangle_1$), with
$\alpha_j$ $\!=$ $\!\alpha_j(x)$ being the complex amplitude and
$r$ $\!=$ $\!r(x)$ being the purity parameter. To see this, consider
Alice's and Bob's signal pulses which, after leaving $\hat{K}_{02}$ and
$\hat{K}_{13}$, enter the transmission line at $x$ $\!=$ $\!0$. Their state
$\hat{\varrho}(0)$ $\!=$ $\!|\Psi\rangle_{01}\langle\Psi|$, with
$|\Psi\rangle_{01}$ given Eq.~(\ref{st}), can be written in the form of
Eq.~(\ref{initstate}) with
\begin{subequations}
\begin{eqnarray}
  r(0)&=&1,
  \\
  \alpha_j(0)&=&\alpha.
\end{eqnarray}
\label{initvalue-a}
\end{subequations}
A solution of Eq.~(\ref{me}) obeying the initial condition
Eqs.~(\ref{initvalue-a}) is again given by a state of the form
Eq.~(\ref{initstate}) with
\begin{subequations}
\begin{eqnarray}
  r(x)&=&\mathrm{e}^{-2\left(2|\alpha|^2-|\alpha_0|^2-|\alpha_1|^2\right)},
  \label{sa}\\
  \alpha_j(x)&=&T_j^{\frac{x}{L}}\alpha.
  \label{sb}
\end{eqnarray}
\label{solution-a}
\end{subequations}
By inserting Eq.~(\ref{sb}) into Eq.~(\ref{sa}) we see that for large $|\alpha|$,
the purity parameter $r$ decreases much faster then the amplitude $|\alpha|$. If
the length of the transmission line $l$ is small compared to the characteristic
length $L$ of transparency, we may therefore neglect the damping of the
amplitude and the parameters of the state at the end of the transmission line
can be approximated by
\begin{subequations}
\begin{eqnarray}
  r(l)&\stackrel{l\ll L}{\approx}&
  \exp\left[4|\alpha|^2\left(\ln T_{0}+\ln T_{1}\right)l/L\right],
  \\
  \alpha_j(l)&\stackrel{l\ll L}{\approx}&\alpha.
\end{eqnarray}
\label{approxsolution-a}
\end{subequations}
Note that the sensitivity to decoherence is due to the fact that the entangled
states are amplified to macroscopic level. The amplitude plays here the role of
the separating parameter as does the spatial separation in case of two
particles. The information contained in the entanglement itself (see below) is
not inreased by Alice's and Bob's local operations and does not exceed one ebit
\cite{catEnk} as originally prepared in Eq.~(\ref{s}). While the limit
$|\alpha|$ $\!\rightarrow$ $\!\infty$ represents an unstable state, comparable
to a bowl sitting on a sphere, that cannot be perpetuated, the type invariance
of the state Eq.~(\ref{initstate}) under the influence of transmission loss makes
it a candidate for a carrier of quantum information in the mesoscopic regime.
\section{\label{sec3}
       State purification}
\subsection{\label{sec3.1}
         Principle}
At the receiving end of the transmission line, there is another pair of
stations run by, say, Claire and Denis, who receive a state $\hat{\varrho}(l)$
given by Eq.~(\ref{initstate}) together with Eqs.~(\ref{approxsolution-a}).
Their task is to rebuild the original pure Schr\"odinger cat-like entangled
state $\hat{\varrho}(0)$ as given in Eq.~(\ref{st}). If Claire and Denis can
communicate with a third party by means of classical signals only, they must
distill $\hat{\varrho}(0)$ from a number of preparations $\hat{\varrho}(l)$.

Since we have assumed that the change of the complex amplitude due to
scattering loss during transmission can be neglected, compare
Eqs.~(\ref{approxsolution-a}), the only complex amplitudes occuring in the
signal state are $\pm\alpha$. On the other hand, for sufficiently small
$\exp(-|\alpha|^2)$, the coherent states $|+\alpha\rangle$ and
$|\!-\!\alpha\rangle$ become orthogonal, and therefore form an orthonormal
basis, which we denote by $\uparrow$ corresponding to $\alpha$ and $\downarrow$
corresponding to $-\alpha$. In this way, the signal pulses can be regarded as a
bipartite quantum system. The signal state $\hat{\varrho}(l)$ at the exit of the
transmission line can then be rewritten as a mixture
\begin{eqnarray}
  \hat{\varrho}(l)
  &=&\frac{|\uparrow\downarrow\rangle\langle\uparrow\downarrow|
  +|\downarrow\uparrow\rangle\langle\downarrow\uparrow|}{2}
  +r\frac{|\uparrow\downarrow\rangle\langle\downarrow\uparrow|
  +|\downarrow\uparrow\rangle\langle\uparrow\downarrow|}{2}
  \nonumber\\
  &=&\frac{1+r}{2}|\Psi^+\rangle\langle\Psi^+|
  +\frac{1-r}{2}|\Psi^-\rangle\langle\Psi^-|
\label{bellmix1}
\end{eqnarray}
of the Bell states
\begin{equation}
  |\Psi^\pm\rangle
  =\frac{|\uparrow\downarrow\rangle\pm|\downarrow\uparrow\rangle}{\sqrt{2}}.
\end{equation}
In \cite{entangledBennett}, the entanglement of formation $E$ of a state
$\hat{\varrho}$ of a bipartite quantum system is defined and discussed. For a
general mixture of Bell states, the expression
\begin{equation}
  E=-x_+\log_2x_+-x_-\log_2x_-
\label{E(r)}
\end{equation}
is obtained, where $x_\pm$ $\!=$ $\!1/2\pm\sqrt{p(1-p)}$, with $p$ being the
maximum of $1/2$ and the eigenvalues of $\hat{\varrho}$. For the mixture
Eq.~(\ref{bellmix1}), this is $p$ $\!=$ $\!(1+|r|)/2$. A plot of the entanglement
Eq.~(\ref{E(r)}) as a function of $|r|$ of the state Eq.~(\ref{bellmix1}) in
Fig.~\ref{fig2} (solid line) reveals that in the case considered here,
entanglement and purity can be regarded as synonyms. In this sense, the state
purification and detection represent an entanglement purification and detection.

\begin{figure}[ht]
\includegraphics[width=8.6cm]{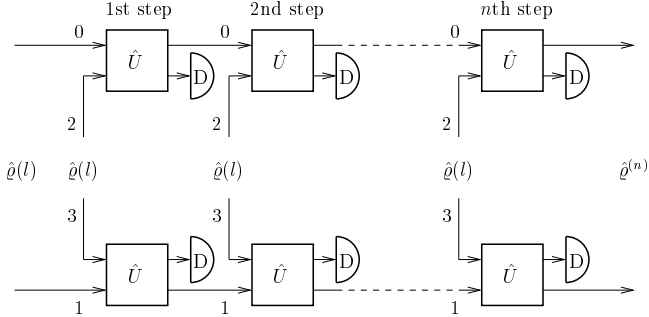}
\caption{\label{fig7}
Principle of random purification of a mixture Eq.~(\ref{bellmix1}). Depending
on the occurrence of $\uparrow$ and $\downarrow$ detections, the output state
$\hat{\varrho}^{(n)}$ becomes $|\Psi^\pm\rangle$ with probability
$(1\pm r)/2$ for sufficiently large (even) $n$.
}
\end{figure}

\begin{figure}[ht]
\includegraphics[width=6.9cm]{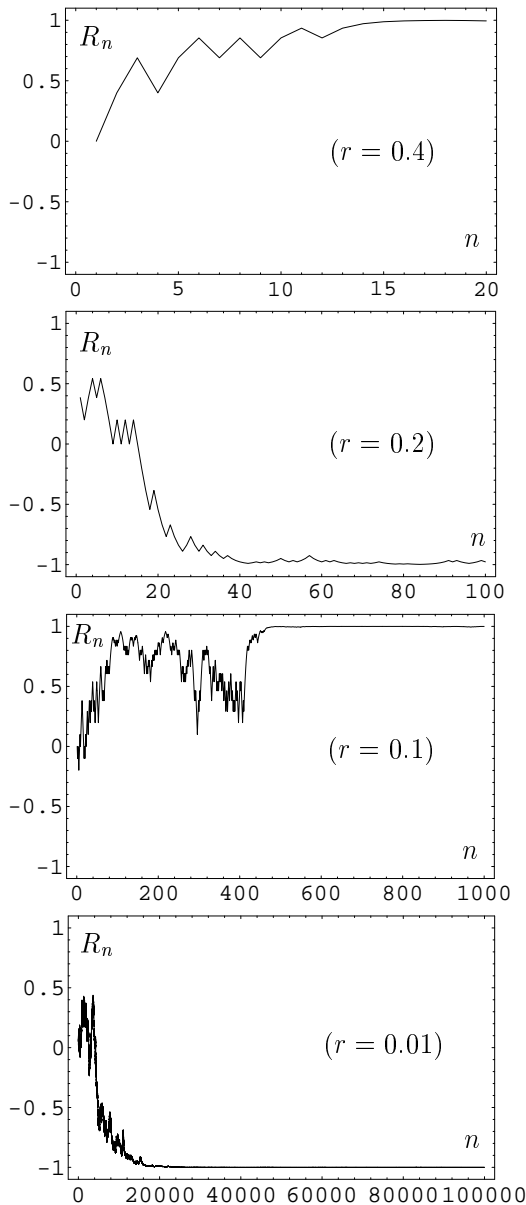}
\caption{\label{fig6}
Random walk of the purity parameter $R_n$ according to Eqs.~(\ref{rec-a}) seen
on different scales appropriate for the respective $r$. This behavior is
obtained if the scheme shown in Fig.~\ref{fig7} is run without intervening
manually.
}
\end{figure}

Let us explain the method of purification we are going to apply before
considering its physical implementation. Detailed work on general properties
of purification protocols is done in \cite{entangledBennett}. Fig.~\ref{fig7}
shows the principle.
Each unit $\hat{U}$ describes a two-qubit quantum gate whose action is defined
by
\begin{equation}
  \hat{U}
  \left(
  \begin{array}{rrrr}
  |\downarrow\downarrow\rangle \\
  |\downarrow\uparrow\rangle \\
  |\uparrow\downarrow\rangle \\
  |\uparrow\uparrow\rangle
  \end{array}
  \right)
  =
  \frac{1}{\sqrt{2}}\left(
  \begin{array}{rrrr}
  1 & 1 & 0 & 0 \\
  0 & 0 & 1 & -1 \\
  0 & 0 & 1 & 1 \\
  1 & -1 & 0 & 0
  \end{array}
  \right)
  \left(
  \begin{array}{rrrr}
  |\downarrow\downarrow\rangle \\
  |\downarrow\uparrow\rangle \\
  |\uparrow\downarrow\rangle \\
  |\uparrow\uparrow\rangle
  \end{array}
  \right).
\label{gate}
\end{equation}
The detection devices D at the output modes 2 and 3 of these gates detect
either a state $|\downarrow\rangle$ or $|\uparrow\rangle$. Initially, the two
signal modes 0 and 1 are prepared in a state Eq.~(\ref{bellmix1}). The same holds
for the input modes 2 and 3 of all gates. The output state $\hat{\varrho}^{(n)}$
of the signal modes 0 and 1 after the $n$th step can be obtained by applying
Eq.~(\ref{gate}) to Eq.~(\ref{bellmix1}) after straightforward algebra.
It is given by
\begin{subequations}
\begin{eqnarray}
  \hat{\varrho}^{(n=2k)}
  =\frac{1+R_n}{2}|\Psi^+\rangle\langle\Psi^+|
  +\frac{1-R_n}{2}|\Psi^-\rangle\langle\Psi^-|,\quad&&
  \\
  \hat{\varrho}^{(n=2k+1)}
  =\frac{1+R_n}{2}|\Phi^+\rangle\langle\Phi^+|
  +\frac{1-R_n}{2}|\Phi^-\rangle\langle\Phi^-|.\quad&&
\end{eqnarray}
\label{nthstates-a}
\end{subequations}
The switch from $|\Psi^\pm\rangle$ to the Bell states
\begin{equation}
  |\Phi^\pm\rangle
  =\frac{|\uparrow\uparrow\rangle\pm|\downarrow\downarrow\rangle}{\sqrt{2}}.
\end{equation}
in case of odd $n$ is not relevant for our considerations since the
$|\Psi^\pm\rangle$ may be reobtained by applying a single-qubit gate in mode 0
or 1. The purity parameter follows from the recursion
\begin{subequations}
\begin{eqnarray}
  R_n&=&\frac{R_{n-1}\pm r}{1\pm rR_{n-1}},\\
  R_0&=&r.
\end{eqnarray}
\label{rec-a}
\end{subequations}
A plus sign has to be used if the same state is detected in mode 2 and 3 during
the $n$th step, and a minus sign if different states are detected. The
respective probabilities are
\begin{equation}
  p_\pm=\frac{1\pm rR_{n-1}}{2}.
\label{ppm}
\end{equation}
We see that if $R_{n-1}$ $\!=$ $\!\pm1$ then $R_n$ $\!=$ $\!\pm1$. That is, a pure
signal state remains unchanged, independently of $r$. On the other hand, if
$r$ $\!=$ $\!0$ then $R_n$ $\!=$ $\!R_{n-1}$, i.e., the setup cannot alter the
signal state without \textquotedblleft{resources}\textquotedblright\,. Furthermore, if
$|R_{n-1}|$ $\!<$ $\!1$ and assuming that $|r|$ $\!<$ $\!1$, we obtain
$|R_n|$ $\!<$ $\!1$, i.e., a mixed signal state cannot be purified in a
finite number of steps. Assume now that we don't know the measurement results.
The identity
\begin{equation}
  p_+\left(\frac{R_{n-1}+r}{1+rR_{n-1}}\right)
  + p_-\left(\frac{R_{n-1}-r}{1-rR_{n-1}}\right) = R_{n-1}
\end{equation}
reveals that the expectation value of the purity parameter is not altered from
step to step and therefore given by the initial value r. The change of
variance during the $n$th step is the average of the term
\begin{eqnarray}
  &&p_+\left(\frac{R_{n-1}+r}{1+rR_{n-1}}-r\right)^2
  +p_-\left(\frac{R_{n-1}-r}{1-rR_{n-1}}-r\right)^2
  \nonumber\\
  &&-(R_{n-1}-r)^2
  \nonumber\\
  &=&\frac{r^2(1-R_{n-1}^2)^2}{1-r^2R_{n-1}^2}
\end{eqnarray}
over the distribution of $R_{n-1}$. We see that this change is positive unless
the distribution of $R_{n-1}^2$ is sharp with $R_{n-1}^2$ $\!=$ $\!1$. The
probability distribution of a random variable $R_\infty$ $\!\in$ $\![-1,1]$ with
a given expectation value $r$ $\!\in$ $\![-1,1]$ that maximises variance however
is $R_\infty$ $\!=$ $\!\pm1$ with probability $(1\pm r)/2$. This is just the
asymptotic probability distribution of the purity parameter $R_n$ for
$n$ $\!\rightarrow$ $\!\infty$ obtained by running the scheme Fig.~\ref{fig7}
without knowing or performing detections. If the detection results are known,
the progression of the $R_n$ [and with it the signal state
$\hat{\varrho}^{(n)}$] can be computed from Eqs.~(\ref{rec-a}). The random walk
of the purity parameter resulting from a monitored free run of the scheme
Fig.~\ref{fig7} is simulated in Fig.~\ref{fig6}.
As Fig.~\ref{fig6} illustrates, $R_\infty$ $\!=$ $\!\pm1$ with probability
$(1\pm r)/2$. If the initial value $r$ is unknown, the output state will be
unknown, but the progress of the purification can still be monitored on the
basis of the joint probabilities Eq.~(\ref{ppm}). The purification is completed
as soon as they no longer fluctuate. A successful purification is identified by
the inequality $p_+$ $\!>$ $\!p_-$.

By repeatedly running the simulation Fig.~\ref{fig6} one may estimate the
average number $\bar{n}$ of steps until $1-|R_n|$ has fallen below a given
$\varepsilon$ $\!\ll$ $\!1$. For instance, $\varepsilon$ $\!=$ $\!10^{-5}$ gives
\begin{displaymath}
  \begin{array}{r} r \\ \bar{n} \end{array}
  \qquad
  \begin{array}{rrrrrrrrr}
  0.9 &\; 0.8 &\; 0.7 &\; 0.6 &\; 0.5 &\; 0.4 &\; 0.3 &\; 0.2 &\; 0.1 \\
    5 &\;   7 &\;  10 &\;  14 &\;  23 &\;  37 &\;  66 &\; 154 &\; 609
\end{array}
\end{displaymath}
depending on $r$, which tells the average number of steps needed until the
variance of the $R_n$-distribution remains constant and purification is
completed.

An observer unaware of the measurement outcome of the $n$th step observes the
average
\begin{equation}
  p_+\hat{\varrho}^{(n)}_{|+}+ p_-\hat{\varrho}^{(n)}_{|-}
  =\hat{\varrho}^{(n-1)}_{(\Psi\leftrightarrow\Phi)}
\end{equation}
of the states $\hat{\varrho}^{(n)}_{|\pm}$ resulting from events $\pm$ with
probabilities $p_\pm$, compare Eqs.~(\ref{rec-a}) and Eq.~(\ref{ppm}), which is,
apart from the Bell state flip, just the state before this step. As a
consequence, such an observer perceives states according to
Eqs.~(\ref{nthstates-a}) but with Eqs.~(\ref{rec-a}) replaced with
$R_n$ $\!=$ $\!R_0$. In this sense, the scheme in Fig.~\ref{fig7} represents a
\textquotedblleft{quantum state guide}\textquotedblright\, for an arbitrary mixture of
either $|\Psi^\pm\rangle$ or $|\Phi^\pm\rangle$.
\subsection{\label{sec3.2}
         Implementation}
\begin{figure}[ht]
\includegraphics[width=8.6cm]{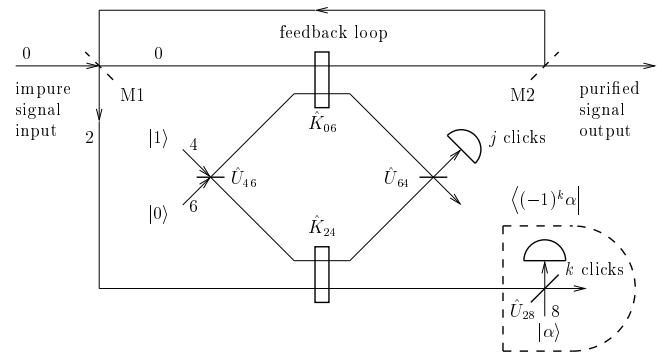}
\caption{\label{fig4}
Purification of entangled mesoscopic coherent states. The pulse whose state is
to be purified  is allowed to enter the loop by lifting mirror M1 where it is
mixed periodically with subsequent signal pulses from which the pure components
are to be distilled. If the desired purity is reached, the pulse is released by
lifting mirror M2.
}
\end{figure}

Let us consider, e.g., Claire's equipment. In order to implement the operation
Eq.~(\ref{gate}) on coherent states, she may be equipped with a device shown in
Fig.~\ref{fig4}.
The scheme consists of mirrors M1 and M2 removable on demand and two cross-Kerr
couplers $\hat{K}_{06}$ and $\hat{K}_{24}$, themselves coupled to each other by
a Mach-Zehnder interferometer. The latter consists of balanced beam splitters
$\hat{U}_{46}(T=-R=1/\sqrt{2})$ and $\hat{U}_{64}(T=R=1/\sqrt{2})$ and is fed
with a pulse in a single-photon state $|1\rangle$. (Alternatively, a coherent
state may be used instead.) After passing the cross-Kerr couplers, this photon
is detected either in mode 4 or mode 6, depending on whether the
ON/OFF-detector gives a signal or not. (An ON/OFF-detector is a photodetector
able to discriminate between absence and presence of photons. In Fig.~\ref{fig4}
this is denoted by $j,k$ $\!=$ 0 and 1 \textquotedblleft{clicks}\textquotedblright\,,
respectively.) The purpose of the detection device drawn in dashed lines is to
discriminate between coherent states $|\!\pm\!\alpha\rangle$. In our
application, the reduced state at its input port will be a mixture of coherent
states $|\!\pm\!\alpha\rangle$, therefore this discrimation can be achieved by
mixing the input with a pulse in a coherent state $|\alpha\rangle$ using beam
splitter $\hat{U}_{28}(T=R=1/\sqrt{2})$ and detecting photon presence with an
ON/OFF-detector in mode 8. The detector can only give a signal if the sign of
the coherent state at the input is negative.

To see how the scheme works, we remove mirrors M1 and M2, and assume that the
input mode 2 is prepared in a coherent state $|\!\pm\!\alpha\rangle$. If the
ON/OFF-detectors give $j$ and $k$ clicks ($j,k$ $\!=$ $\!0,1$) as shown in the
figure, the action of the whole setup on the signal mode 0 can be described by
a conditional single-mode operator
\begin{eqnarray}
  &&\hat{Y}_0(\pm\alpha|j,k)
  \nonumber\\
  &&=\,_2\langle(-1)^k\alpha|\,_4\langle j|\,_6\langle1\!-\!j|
  \hat{U}|1\rangle_4|0\rangle_6|\!\pm\!\alpha\rangle_2
  \nonumber\\
  &&=\frac{1}{2}(-1)^{k(\hat{n}_0+j)}\mathrm{e}^{-|\alpha|^2}
  \Bigl[\mathrm{e}^{\pm|\alpha|^2}(-1)^{\hat{n}_0
  +j}+\mathrm{e}^{\mp|\alpha|^2}\Bigr]
  \nonumber\\
  &&\approx\frac{1}{2}(-1)^{k(\hat{n}_0+j)}(\mp1)^{\hat{n}_0+j}
  \quad(\mathrm{e}^{-|\alpha|^2}\ll1),
\label{purY}
\end{eqnarray}
where $\hat{U}$ $\!=$ $\!\hat{U}_{64}\hat{K}_{24}\hat{K}_{06}\hat{U}_{46}$.
Assume now that Claire receives two consecutive pulses from Alice, each
corresponding to a state of the form $\hat{\varrho}(l)$ given by
Eq.~(\ref{initstate}) together with Eqs.~(\ref{approxsolution-a}). Let us further
assume that the purity parameter of the first is $R$ [we denote this state by
$\hat{\varrho}_{01}(R)$] and the second $r$. Claire feeds the first pulse into
$\hat{K}_{06}$ (i.e., M1 removed) and the second into $\hat{K}_{24}$ (i.e., M1
inserted). Denis, who is equipped with an equivalent device described by an
analogous operator $\hat{Y}_1$, does the same with the two pulses he receives
from Bob. The state of the pulses leaving the signal outputs (M2 removed) can
then be written as
\begin{eqnarray}
  \hat{\varrho}^\prime
  &=&\frac{1}{p}\frac{1}{2(1+r\mathrm{e}^{-4|\alpha|^2})}
  \nonumber\\
  &&\times\Bigl[\hat{Y}_0(\alpha)\hat{Y}_1(-\alpha)
  \hat{\varrho}_{01}(R)\hat{Y}_0^\dagger(\alpha)\hat{Y}_1^\dagger(-\alpha)
  \nonumber\\
  &&\quad+\hat{Y}_0(-\alpha)\hat{Y}_1(\alpha)
  \hat{\varrho}_{01}(R)\hat{Y}_0^\dagger(-\alpha)\hat{Y}_1^\dagger(\alpha)
  \nonumber\\
  &&\quad+r\hat{Y}_0(\alpha)\hat{Y}_1(-\alpha)
  \hat{\varrho}_{01}(R)\hat{Y}_0^\dagger(-\alpha)\hat{Y}_1^\dagger(\alpha)
  \nonumber\\
  &&\quad+r\hat{Y}_0(-\alpha)\hat{Y}_1(\alpha)
  \hat{\varrho}_{01}(R)\hat{Y}_0^\dagger(\alpha)\hat{Y}_1^\dagger(-\alpha)\Bigr],
\label{purst}
\end{eqnarray}
where $\hat{Y}_l(\pm\alpha)$ $\!=$ $\!\hat{Y}_l(\pm\alpha|j_l,k_l)$.
By inserting the approximated expression of Eq.~(\ref{purY}) into
Eq.~(\ref{purst}), we obtain
\begin{eqnarray}
  \hat{\varrho}^\prime
  &=&\frac{1}{2^4p}
  \frac{(1+R^\prime\mathrm{e}^{-4|\alpha|^2})}{(1+r\mathrm{e}^{-4|\alpha|^2})}
  \frac{[1+(-1)^{j_0+j_1}rR]}{(1+R\mathrm{e}^{-4|\alpha|^2})}
  \nonumber\\
  &&\times\hat{P}\hat{\varrho}_{01}(R^\prime)\hat{P}^\dagger.
\label{rhoprime}
\end{eqnarray}
Here,
\begin{equation}
  \hat{P}=(-1)^{k_0\hat{n}_0+(1-k_1)\hat{n}_1}
\end{equation}
is a product of phase shifts which Claire and Denis may compensate by inserting
appropriate phase plates into their signal modes. In this way, a signal output
state $\hat{\varrho}_{01}(R^\prime)$ can be prepared that has the same form as
the input state $\hat{\varrho}_{01}(R)$ except for the purity parameter which
has changed to
\begin{equation}
  R^\prime=\frac{R+(-1)^{j_0+j_1}r}{1+(-1)^{j_0+j_1}rR}.
\label{rprime}
\end{equation}
From Eq.~(\ref{rhoprime}) we obtain the probability
\begin{equation}
  p(j_0,k_0,j_1,k_1)\approx\frac{1+(-1)^{j_0+j_1}rR}{2^4}
  \quad(\mathrm{e}^{-|\alpha|^2}\ll1).
\label{jp}
\end{equation}
Claire and Denis may now insert a mirror M2 after the first run to feed the
output pulse with purity parameter $R^\prime$ back into the input port while at
the same time a new signal pulse from the transmission line with purity
parameter $r$ enters the other input port via M1. In this way, the purity
parameter of the signal pulse circulating in the feedback loop changes
stepwise, e.g., after having completed its $n$th round trip according to
Eqs.~(\ref{rec-a}) with probability Eq.~(\ref{ppm}). A positive (negative) sign
is realized if $j_0$ $\!+$ $\!j_1$ is even (odd). Claire's and Denis's only
remaining task is to transmit the measurement results $j$ and $k$ to a third
party where they are collected. In those cases when $R$ $\!\rightarrow$ $\!1$,
the third party answers them with the instruction to open their mirrors M2.
\begin{itemize}
\item \textit{Alternative:}
\end{itemize}
If Claire and Denis have the option to send a light pulse to a third party,
they may proceed as in the alternative explained in Sec.~\ref{sec2}. The action
of the conditional operator Eq.~(\ref{ytilde}) on the signal state
$\hat{\varrho}(l)$ then yields
\begin{equation}
  \frac{1}{p}\hat{\tilde{Y}}\hat{\varrho}(l)\hat{\tilde{Y}}^\dagger
  =(-1)^{\hat{n}_1}\hat{\varrho}(0)(-1)^{\hat{n}_1},
\end{equation}
i.e., after an additional phase shift $(-1)^{\hat{n}_1}$, the original pure
state is reobtained in an instant without need for additional signal pulses.
The success probability becomes
\begin{eqnarray}
  p&=&\frac{|\langle0|\beta\rangle\langle1|\beta\rangle|^2
  (1+r)(1+\mathrm{e}^{-4|\alpha|^2})}
  {1+r\mathrm{e}^{-4|\alpha|^2}}
  \nonumber\\
  &=&4p_\mathrm{max}\frac{1+r}{1+r\mathrm{e}^{-4|\alpha|^2}}
  \quad(|\beta|^2=1/2),
\end{eqnarray}
compare Eq.~(\ref{pmax}).
\section{\label{sec4}
       State detection}
We return to Fig.~\ref{fig1}. Assume that Claire and Denis receive a sequence of
signal pulses sent by Alice and Bob via the lossy transmission line, each in a
state $\hat{\varrho}(l)$ given by Eq.~(\ref{initstate}) together with
Eqs.~(\ref{approxsolution-a}). Their task is now to measure the purity parameter
$r$ which is assumed to be completely unknown. Neither Claire nor Denis is able
to perform this measurement alone, since their reduced signal state becomes
according to
\begin{eqnarray}
  \mathrm{Tr}_{0/1}\hat{\varrho}(l)
  &=&\frac{1}{2(1+r\mathrm{e}^{-4|\alpha|^2})}
  \Bigl[(|\alpha\rangle\langle\alpha|+|\!-\!\alpha\rangle\langle-\alpha|)
  \nonumber\\
  &&+r\mathrm{e}^{-2|\alpha|^2}
  (|\alpha\rangle\langle-\alpha|+|\!-\!\alpha\rangle\langle\alpha|)\Bigr]
\label{reduced}
\end{eqnarray}
a mixture of $|\alpha\rangle$ and $|\!-\!\alpha\rangle$ for small
$\exp(-|\alpha|^2)$. Claire and Denis can however insert a photodetector into
their signal mode and send a classical bit to a third party depending on
whether they have detected an even (event $e$) or odd (event $u$) number of
photons. After a certain number of trials, the third party estimates the
coincidence rate
\begin{equation}
  M=p(e,e)+p(u,u)-p(e,u)-p(u,e)
\end{equation}
of even and odd events in Claire's and Denis's measurement.
Inserting the joint probability
\begin{eqnarray}
  p(m,n)&=&\,_0\langle m|\,_1\langle n|\hat{\varrho}(l)|m\rangle_0|n\rangle_1
  \nonumber\\
  &=&\frac{|\langle m|\alpha\rangle\langle n|\alpha\rangle|^2[1+r(-1)^{m+n}]}
  {1+r\mathrm{e}^{-4|\alpha|^2}}
\end{eqnarray}
of detecting $m$ photons in mode 0 and $n$ photons in mode 1 gives
\begin{eqnarray}
  M&=&\sum_{m,n=0}^\infty\Bigl[p(2m,2n)+p(2m+1,2n+1)
  \nonumber\\
  &&\qquad-p(2m,2n+1)-p(2m+1,2n)\Bigr]
  \nonumber\\
  &=&\frac{r+\mathrm{e}^{-4|\alpha|^2}}{1+r\mathrm{e}^{-4|\alpha|^2}}
  \;\approx\;r\quad(\mathrm{e}^{-|\alpha|^2}\ll1).
\label{coin}
\end{eqnarray}
In this way, the purity parameter can be measured as coincidence rate. This
possibility is however cumbersome and difficult to implement since a
discrimination between even and odd photon numbers requires single-photon
resolution of the photodetectors for $|\alpha|\gg1$.
\begin{itemize}
\item \textit{Alternative:}
\end{itemize}
It is however possible to measure $r$ nondestructively at a separate location.
To achieve this, Claire and Denis prepare modes 6 and 7 in coherent states
$|\gamma\rangle$. After coupling them to the signal modes by cross-Kerr
couplers $\hat{K}_{06}$ and $\hat{K}_{17}$, these auxiliary modes are
recombined in a double-slit setup as shown in Fig.~\ref{fig1}. The purity
parameter $r$ can there be observed directly as the contrast of the
interference fringes emerging on the screen. To see this, we first define the
operator
\begin{equation}
  \hat{I}
  =\hat{U}^\dagger(\hat{a}_{6}+\hat{a}_{7})^\dagger(\hat{a}_{6}+\hat{a}_{7})\hat{U},
\label{int}
\end{equation}
where
\begin{equation}
  \hat{U}=\mathrm{e}^{-\mathrm{i}(\varphi_{6}\hat{n}_{6}
  +\varphi_{7}\hat{n}_{7})}
\end{equation}
depends on the phases $\varphi_{6}$ and $\varphi_{7}$ corresponding to the
optical distances between the given point on the screen and aperture 6 or 7.
The observed light intensity at some given point on the screen is then
approximately given by the expectation value
\begin{equation}
  I\sim\langle\hat{I}\rangle
  =\mathrm{Tr}_{67}(\hat{\varrho}_{67}\hat{I}),
\end{equation}
compare \cite{slitWeb}. Here, $\hat{\varrho}_{67}$ is the state of the two modes
corresponding to the two input ports (slits or pinholes). While the spatial
distribution of the interference pattern is determined by the geometry of the
setup [and the response of the photographic medium or eye which is assumed to
be linear in Eq.~(\ref{int})], its contrast is determined by the input state.
Inserting
\begin{equation}
  \hat{\varrho}_{67}
  =\mathrm{Tr}_{01}\left(\hat{K}_{17}\hat{K}_{06}
  |\gamma\rangle_{6}|\gamma\rangle_{7}
  \,\hat{\varrho}(l)\,
  \,_{6}\langle\gamma|\,_{7}\langle\gamma|
  \hat{K}_{06}^\dagger\hat{K}_{17}^\dagger\right)
\label{varrho67}
\end{equation}
yields
\begin{equation}
  \langle\hat{I}\rangle=2|\gamma|^2\left(1+\frac{r+\mathrm{e}^{-4|\alpha|^2}}
  {1+r\mathrm{e}^{-4|\alpha|^2}}\cos\Delta\varphi\right),
\label{intensity}
\end{equation}
whose maxima and minima
\begin{equation}
  I_{^\mathrm{max}_\mathrm{min}}\sim2|\gamma|^2(1\pm|M|),
\label{imaxmin}
\end{equation}
compare Eq.~(\ref{coin}), are located at those of the phase difference
$\Delta\varphi=\varphi_{7}-\varphi_{6}$. The contrast (visibility) is defined by
\begin{equation}
  \frac{I_\mathrm{max}-I_\mathrm{min}}{I_\mathrm{max}+I_\mathrm{min}}=|M|
  \approx|r|\quad(\mathrm{e}^{-|\alpha|^2}\ll1).
\end{equation}
The sign of $r$ follows from the intensity at the
symmetry center of the setup for which $\Delta\varphi$ $\!=$ $\!0$. If
$r$ $\!>$ $\!0$, $I_\mathrm{max}$ is observed, otherwise $I_\mathrm{min}$.
In this way, the contrast of the interference seen on the screen provides a
direct \textquotedblleft{naked-eye}\textquotedblright\, estimation of the purity
and entanglement. Note that since $r$ is the only unknown parameter, its
determination here amounts to a complete knowledge of the state
$\hat{\varrho}(l)$.

To estimate the measurement accuracy of the interference contrast, we
consider the variance of $\hat{I}$. By applying Eq.~(\ref{varrho67}) we obtain
\begin{eqnarray}
  \sigma^2I
  &\sim&\langle\hat{I}^2\rangle-\langle\hat{I}\rangle^2
  \\
  &&=4|\gamma|^4(\cos^2\Delta\varphi-1)
  +2(1+2|\gamma|^2)\langle\hat{I}\rangle-\langle\hat{I}\rangle^2,
  \nonumber
\end{eqnarray}
and inserting Eq.~(\ref{imaxmin}) gives the relative uncertainties
\begin{equation}
  \frac{\sigma^2I_{^\mathrm{max}_\mathrm{min}}}
  {I_{^\mathrm{max}_\mathrm{min}}^2}
  =\frac{1+2|\gamma|^2}{|\gamma|^2(1\pm|M|)}-1
\label{reluncertI}
\end{equation}
of the intensity extremals. We now use these to estimate the
relative uncertainty
\begin{eqnarray}
  \frac{\sigma^2|M|}{|M|^2}
  &=&\frac{\left(\frac{\partial|M|}{\partial I_\mathrm{max}}\right)^2
  \sigma^2I_\mathrm{max}
  +\left(\frac{\partial|M|}{\partial I_\mathrm{min}}\right)^2
  \sigma^2I_\mathrm{min}}{|M|^2}
  \nonumber\\
  &=&\left(
  \frac{2I_\mathrm{max}I_\mathrm{min}}{I_\mathrm{max}^2-I_\mathrm{min}^2}
  \right)^2\left(
  \frac{\sigma^2I_\mathrm{max}}{I_\mathrm{max}^2}
  +\frac{\sigma^2I_\mathrm{min}}{I_\mathrm{min}^2}
  \right)
\label{reluncertM}
\end{eqnarray}
of the contrast. After inserting Eq.~(\ref{reluncertI}) together with
Eq.~(\ref{imaxmin}), Eq.~(\ref{reluncertM}) becomes
\begin{eqnarray}
  \frac{\sigma^2|M|}{|M|^2}
  &=&\frac{|M|^{-2}-1}{2}\left(|\gamma|^{-2}+1+|M|^2\right).
\label{uncert}
\end{eqnarray}
For $|\gamma|$ $\!\rightarrow$ $\!0$ it diverges according to
$(|M|^{-2}\sigma^2|M|)$ $\!\propto$ $\!|\gamma|^{-2}$ as in the case of a
classical interference experiment with coherent states. As a consequence, a
long series of repetitions is required to obtain reliable data.

\begin{figure}[ht]
\includegraphics[width=6.9cm]{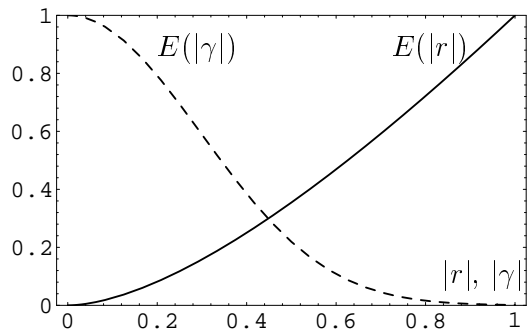}
\caption{\label{fig2}
Entanglement of formation $E(|r|)$ of the signal state at the input ports of
$\hat{K}_{06}$ and $\hat{K}_{17}$ in Fig.~\ref{fig1} as a function of the purity
$|r|$ (solid line).
Entanglement of formation $E(|\gamma|)$ of the signal state at the output ports
of $\hat{K}_{06}$ and $\hat{K}_{17}$ as a function of the amplitude $|\gamma|$
for an initially pure state, $r$ $\!=$ $\!1$ (dashed line).
}
\end{figure}

The question arises, how this nondemolition measurement affects the state of
the signal pulses. The reduced signal state at the output ports of
$\hat{K}_{06}$ and $\hat{K}_{17}$ can be written as
\begin{eqnarray}
  &&\mathrm{Tr}_{67}\left(\hat{K}_{17}\hat{K}_{06}
  |\gamma\rangle_{6}|\gamma\rangle_{7}
  \,\hat{\varrho}(l)\,
  \,_{6}\langle\gamma|\,_{7}\langle\gamma|
  \hat{K}_{06}^\dagger\hat{K}_{17}^\dagger\right)
  \nonumber\\
  &&=C\hat{\varrho}(l)
  +S\Bigl[(-1)^{\hat{n}_1}\hat{\varrho}(l)(-1)^{\hat{n}_1}\Bigr]
  \nonumber\\
  &&\equiv\hat{\varrho}(x>l),
\label{afterstate}
\end{eqnarray}
where $C$ $\!=$ $\!C(2|\gamma|^2)$ and $S$ $\!=$ $\!S(2|\gamma|^2)$,
compare Eqs.~(\ref{cs-a}). 
Eq.~(\ref{afterstate}) reveals that the effect of the measurement on the signal
state $\hat{\varrho}(l)$ can be neglected if $|\gamma|^2$ $\!\ll$ $\!1$,
independently of $|\alpha|$, i.e., the separation of the coherent states. To
analyse the remaining entanglement, we write Eq.~(\ref{afterstate}) in a form
analogous to Eq.~(\ref{bellmix1}),
\begin{eqnarray}
  &&\hat{\varrho}(x>l)=
  \nonumber\\
  &&C\frac{(|\uparrow\downarrow\rangle\langle\uparrow\downarrow|
  +|\downarrow\uparrow\rangle\langle\downarrow\uparrow|)
  +r(|\uparrow\downarrow\rangle\langle\downarrow\uparrow|
  +|\downarrow\uparrow\rangle\langle\uparrow\downarrow|)}{2}
  \nonumber\\
  &&+S\frac{(|\uparrow\uparrow\rangle\langle\uparrow\uparrow|
  +|\downarrow\downarrow\rangle\langle\downarrow\downarrow|)
  +r(|\uparrow\uparrow\rangle\langle\downarrow\downarrow|
  +|\downarrow\downarrow\rangle\langle\uparrow\uparrow|)}{2}
  \nonumber\\
  &&=C\frac{1+r}{2}|\Psi^+\rangle\langle\Psi^+|
  +C\frac{1-r}{2}|\Psi^-\rangle\langle\Psi^-|
  \nonumber\\
  &&\quad+S\frac{1+r}{2}|\Phi^+\rangle\langle\Phi^+|
  +S\frac{1-r}{2}|\Phi^-\rangle\langle\Phi^-|.
\label{bellmix2}
\end{eqnarray}
Eq.~(\ref{bellmix2}) reveals that $\hat{\varrho}(x>l)$ is now a mixture of all
four Bell states. Inserting $p$ $\!=$ $\!C(1+|r|)/2$ into Eq.~(\ref{E(r)}) gives
the entanglement of formation remaining after the nondemolition measurement.
Fig.~\ref{fig2} (dashed line) shows a plot of $E(|\gamma|)$ if the signal was in
a pure state prior to entering the cross-Kerr elements, $r$ $\!=$ $\!1$. The plot
demonstrates that, as a consequence of the high nonlinearity applied, an
amplitude of the order $|\gamma|$ $\!\approx$ $\!1$ suffices to destroy the
entanglement.

On the other hand, a comparison of Eq.~(\ref{bellmix2}) with Eq.~(\ref{uncert})
shows that the prize to pay for a gentle measurement on $\hat{\varrho}(l)$,
i.e., keeping $|\gamma|^2$ $\!\ll$ $\!1$, is a poor accuracy of the obtained
data which has to be compensated by a large number of repetitions.
We may define a relative deviation
\begin{eqnarray}
  \frac{\left\langle\left[\hat{\varrho}(x>l)
  -\hat{\varrho}(l)\right]^2\right\rangle}
  {\left\langle\hat{\varrho}^2(l)\right\rangle}
  &=&(1-C)^2
  \nonumber\\
  &\approx&4|\gamma|^4\quad(|\gamma|\ll1),
\label{statedeviation}
\end{eqnarray}
where the expectation values are evaluated using $\hat{\varrho}(l)$. The product
of Eq.~(\ref{uncert}) and Eq.~(\ref{statedeviation}) vanishes with $|\gamma|^2$
proportional to $|\gamma|^2$.

\begin{figure}[ht]
\includegraphics[width=8.6cm]{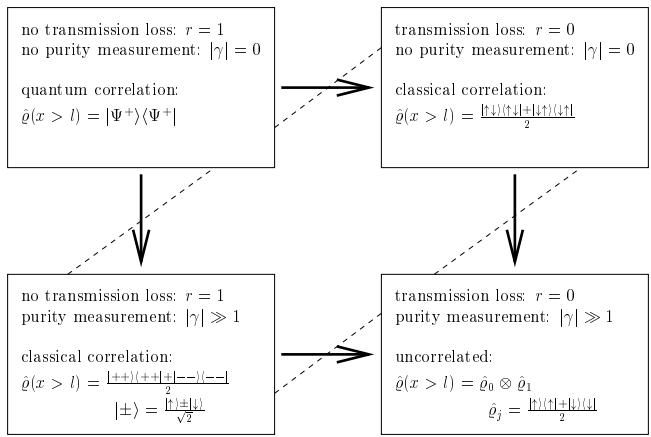}
\caption{\label{fig3}
Complementary behavior of the two loss mechanisms. The scattering along the
transmission line destroys the quantum entanglement by altering the amplitude
and the cross-Kerr elements required for the interference device by altering
the phase. Only their combination leads to an erasure of the remaining
classical correlation.
}
\end{figure}

It may be interesting to note that the complementary behavior of the two types
of interaction between signal modes and environment are reflected in their
effect on entanglement. The lossy transmission channel is assumed to act on the
photon number, leaving the phase unchanged while the nondemolition measurement
acts on the phase, leaving the photon number unchanged. As Fig.~\ref{fig3}
illustrates, each of them alone degrades the correlation from a quantum to
classical level, while their combination is necessary to produce a fully
uncorrelated state.
\section{\label{sec5}
       Conclusion and outlook}
We have considered a superposition of two-mode coherent states with equal
amplitudes but opposite phases under the aspect of preparation, transmission,
purification, and detection.
The states can be prepared in conditional measurement from an entangled
two-mode single-photon state and single-mode coherent states by applying
cross-Kerr elements.
The master equation describing state evolution in a lossy transmission line can
be solved analytically. Its solution shows the well-known transition from a
superposition state to a corresponding mixture.
The original pure state can be extracted from a number of transmitted copies
by local setups which also apply cross-Kerr elements. A monitored run of these
setups leads to a semi-probabilistic self-purifcation of the signal state.
Purity and entanglement of the transmitted state can be regarded as synonyms
and are observable directly as contrast of the interference seen behind a 
double-slit setup. To achieve this, the double-slit setup is fed with
auxilliary modes which were previously coupled to the signal modes by
cross-Kerr elements. A decrease of the signal state perturbation caused by this
nondemolition measurement is connected with an increase of the measurement
uncertainty.

The question arises whether the manipulations discussed in this work can also
be applied to two-mode squeezed vacuum states, which represent according to
\begin{eqnarray}
  |z\rangle_{01}
  &=&\sqrt{1-|z|^2}\,\mathrm{e}^{z\hat{a}_0^\dagger\hat{a}_1^\dagger}|0,0\rangle
  \quad(|z|<1)
  \label{ss}\\
  &=&\,\frac{\sqrt{|z|^{-2}-1}}{\pi}\int\mathrm{d}^2\alpha\,
  \mathrm{e}^{-(|z|^{-1}-1)|\alpha|^2}
  |\alpha,\mathrm{e}^{\mathrm{i}\varphi_z}\alpha^*\rangle
  \nonumber
\end{eqnarray}
continuous superpositions of two-mode coherent states. Analogously to the limit
$|\alpha|$ $\!\rightarrow$ $\!\infty$ in Eq.~(\ref{st}), the transition
$|z|$ $\!\rightarrow$ $\!1$ can be made in Eq.~(\ref{ss}) resulting in an
EPR-like state \cite{eprEinstein,masterWalls,eprBanaszek}, which, in contrast to
Eq.~(\ref{st}), represents a highly entangled state.
\begin{acknowledgments}
This work was supported by the Deutsche Forschungsgemeinschaft.
\end{acknowledgments}

\begin{thebibliography}{17}
\expandafter\ifx\csname natexlab\endcsname\relax\def\natexlab#1{#1}\fi
\expandafter\ifx\csname bibnamefont\endcsname\relax
  \def\bibnamefont#1{#1}\fi
\expandafter\ifx\csname bibfnamefont\endcsname\relax
  \def\bibfnamefont#1{#1}\fi
\expandafter\ifx\csname citenamefont\endcsname\relax
  \def\citenamefont#1{#1}\fi
\expandafter\ifx\csname url\endcsname\relax
  \def\url#1{\texttt{#1}}\fi
\expandafter\ifx\csname urlprefix\endcsname\relax\def\urlprefix{URL }\fi
\providecommand{\bibinfo}[2]{#2}
\providecommand{\eprint}[2][]{\url{#2}}

\bibitem[{\citenamefont{Schr{\"o}dinger}(1935)}]{catSchrodinger}
\bibinfo{author}{\bibfnamefont{E.}~\bibnamefont{Schr{\"o}dinger}},
  \bibinfo{journal}{Naturw.} \textbf{\bibinfo{volume}{23}},
  \bibinfo{pages}{807, 823, 844} (\bibinfo{year}{1935}).

\bibitem[{\citenamefont{Bennett et~al.}(1996)\citenamefont{Bennett, DiVincenzo,
  Smolin, and Wootters}}]{entangledBennett}
\bibinfo{author}{\bibfnamefont{C.~H.} \bibnamefont{Bennett}},
  \bibinfo{author}{\bibfnamefont{D.~P.} \bibnamefont{DiVincenzo}},
  \bibinfo{author}{\bibfnamefont{J.~A.} \bibnamefont{Smolin}},
  \bibnamefont{and} \bibinfo{author}{\bibfnamefont{W.~K.}
  \bibnamefont{Wootters}}, \bibinfo{journal}{Phys. Rev. A}
  \textbf{\bibinfo{volume}{54}}, \bibinfo{pages}{3824} (\bibinfo{year}{1996}).

\bibitem[{\citenamefont{Howell and Yeazell}(2000{\natexlab{a}})}]{catHowell}
\bibinfo{author}{\bibfnamefont{J.~C.} \bibnamefont{Howell}} \bibnamefont{and}
  \bibinfo{author}{\bibfnamefont{J.~A.} \bibnamefont{Yeazell}},
  \bibinfo{journal}{Phys. Rev. A} \textbf{\bibinfo{volume}{62}},
  \bibinfo{pages}{012102} (\bibinfo{year}{2000}{\natexlab{a}}).

\bibitem[{\citenamefont{Gerry}(1999)}]{catGerry}
\bibinfo{author}{\bibfnamefont{C.~C.} \bibnamefont{Gerry}},
  \bibinfo{journal}{Phys. Rev. A} \textbf{\bibinfo{volume}{59}},
  \bibinfo{pages}{4095} (\bibinfo{year}{1999}).

\bibitem[{\citenamefont{Filip}()}]{slitFilip}
\bibinfo{author}{\bibfnamefont{R.}~\bibnamefont{Filip}},
  \eprint{quant-ph/0108119}.

\bibitem[{\citenamefont{Rice and Sanders}(1998)}]{coh1Rice}
\bibinfo{author}{\bibfnamefont{D.~A.} \bibnamefont{Rice}} \bibnamefont{and}
  \bibinfo{author}{\bibfnamefont{B.~C.} \bibnamefont{Sanders}},
  \bibinfo{journal}{Quant. Semiclass. Opt.} \textbf{\bibinfo{volume}{10}},
  \bibinfo{pages}{L41} (\bibinfo{year}{1998}).

\bibitem[{\citenamefont{Sanders and Rice}(1999)}]{coh2Rice}
\bibinfo{author}{\bibfnamefont{B.~C.} \bibnamefont{Sanders}} \bibnamefont{and}
  \bibinfo{author}{\bibfnamefont{D.~A.} \bibnamefont{Rice}},
  \bibinfo{journal}{Opt. Quant. Electron.} \textbf{\bibinfo{volume}{31}},
  \bibinfo{pages}{525} (\bibinfo{year}{1999}).

\bibitem[{\citenamefont{Sanders and Rice}(2000)}]{coh3Rice}
\bibinfo{author}{\bibfnamefont{B.~C.} \bibnamefont{Sanders}} \bibnamefont{and}
  \bibinfo{author}{\bibfnamefont{D.~A.} \bibnamefont{Rice}},
  \bibinfo{journal}{Phys. Rev. A} \textbf{\bibinfo{volume}{61}},
  \bibinfo{pages}{013805} (\bibinfo{year}{2000}).

\bibitem[{\citenamefont{Rice et~al.}(2000)\citenamefont{Rice, Jaeger, and
  Sanders}}]{coh4Rice}
\bibinfo{author}{\bibfnamefont{D.~A.} \bibnamefont{Rice}},
  \bibinfo{author}{\bibfnamefont{G.}~\bibnamefont{Jaeger}}, \bibnamefont{and}
  \bibinfo{author}{\bibfnamefont{B.~C.} \bibnamefont{Sanders}},
  \bibinfo{journal}{Phys. Rev. A} \textbf{\bibinfo{volume}{62}},
  \bibinfo{pages}{012101} (\bibinfo{year}{2000}).

\bibitem[{\citenamefont{Bartlett et~al.}(2001)\citenamefont{Bartlett, Rice,
  Sanders, Daboul, and de~Guise}}]{coh5Rice}
\bibinfo{author}{\bibfnamefont{S.~D.} \bibnamefont{Bartlett}},
  \bibinfo{author}{\bibfnamefont{D.~A.} \bibnamefont{Rice}},
  \bibinfo{author}{\bibfnamefont{B.~C.} \bibnamefont{Sanders}},
  \bibinfo{author}{\bibfnamefont{J.}~\bibnamefont{Daboul}}, \bibnamefont{and}
  \bibinfo{author}{\bibfnamefont{H.}~\bibnamefont{de~Guise}},
  \bibinfo{journal}{Phys. Rev. A} \textbf{\bibinfo{volume}{63}},
  \bibinfo{pages}{042310} (\bibinfo{year}{2001}).

\bibitem[{\citenamefont{Lee and Kim}(2000)}]{teleLee}
\bibinfo{author}{\bibfnamefont{H.-W.} \bibnamefont{Lee}} \bibnamefont{and}
  \bibinfo{author}{\bibfnamefont{J.}~\bibnamefont{Kim}},
  \bibinfo{journal}{Phys. Rev. A} \textbf{\bibinfo{volume}{63}},
  \bibinfo{pages}{012305} (\bibinfo{year}{2000}).

\bibitem[{\citenamefont{van Enk and Hirota}(2001)}]{catEnk}
\bibinfo{author}{\bibfnamefont{S.~J.} \bibnamefont{van Enk}} \bibnamefont{and}
  \bibinfo{author}{\bibfnamefont{O.}~\bibnamefont{Hirota}},
  \bibinfo{journal}{Phys. Rev. A} \textbf{\bibinfo{volume}{64}},
  \bibinfo{pages}{022313} (\bibinfo{year}{2001}).

\bibitem[{\citenamefont{Howell and
  Yeazell}(2000{\natexlab{b}})}]{triggerHowell}
\bibinfo{author}{\bibfnamefont{J.~C.} \bibnamefont{Howell}} \bibnamefont{and}
  \bibinfo{author}{\bibfnamefont{J.~A.} \bibnamefont{Yeazell}},
  \bibinfo{journal}{Phys. Rev. A} \textbf{\bibinfo{volume}{62}},
  \bibinfo{pages}{032311} (\bibinfo{year}{2000}{\natexlab{b}}).

\bibitem[{\citenamefont{Schubert and Weber}(1993)}]{slitWeb}
\bibinfo{author}{\bibfnamefont{M.}~\bibnamefont{Schubert}} \bibnamefont{and}
  \bibinfo{author}{\bibfnamefont{G.}~\bibnamefont{Weber}},
  \emph{\bibinfo{title}{Quantentheorie}} (\bibinfo{publisher}{Spektrum Akad.
  Verlag}, \bibinfo{address}{Heidelberg}, \bibinfo{year}{1993}).

\bibitem[{\citenamefont{Walls and Milburn}(1994)}]{masterWalls}
\bibinfo{author}{\bibfnamefont{D.~F.} \bibnamefont{Walls}} \bibnamefont{and}
  \bibinfo{author}{\bibfnamefont{G.~J.} \bibnamefont{Milburn}},
  \emph{\bibinfo{title}{Quantum Optics}} (\bibinfo{publisher}{Springer-Verlag},
  \bibinfo{address}{Berlin}, \bibinfo{year}{1994}).

\bibitem[{\citenamefont{Einstein et~al.}(1935)\citenamefont{Einstein, Podolsky,
  and Rosen}}]{eprEinstein}
\bibinfo{author}{\bibfnamefont{A.}~\bibnamefont{Einstein}},
  \bibinfo{author}{\bibfnamefont{B.}~\bibnamefont{Podolsky}}, \bibnamefont{and}
  \bibinfo{author}{\bibfnamefont{N.}~\bibnamefont{Rosen}},
  \bibinfo{journal}{Phys. Rev.} \textbf{\bibinfo{volume}{47}},
  \bibinfo{pages}{777} (\bibinfo{year}{1935}).

\bibitem[{\citenamefont{Banaszek and W{\'{o}}dkiewicz}(1999)}]{eprBanaszek}
\bibinfo{author}{\bibfnamefont{K.}~\bibnamefont{Banaszek}} \bibnamefont{and}
  \bibinfo{author}{\bibfnamefont{K.}~\bibnamefont{W{\'{o}}dkiewicz}},
  \bibinfo{journal}{Act. Phys. Slov.} \textbf{\bibinfo{volume}{49}},
  \bibinfo{pages}{491} (\bibinfo{year}{1999}).

\end{thebibliography}

\end{document}